\documentclass{iopjournal}

\newcommand{\C}{$^\circ$C}
\newcommand{\htime}[1]{ \rule[2.5pt]{0.25cm}{0.4pt}\,(#1\,h)$\rightarrow$ }

\begin{document}

\articletype{Paper}

\title
{Quasi-linear `non-metallic' resistivity 
  in the distorted-kagome metal CrPdAs}

\author{B.\ Lau$^1$, 
        W.\ Wu$^1$, 
        B.\ Yuan$^2$\orcid{0000-0003-4721-0382},  
        J.\ Nickel$^1$\orcid{0009-0003-5121-8456}, 
     and S.R.\ Julian$^{1,*}$\orcid{0000-0002-5410-8709}} 

\affil{$^1$ Department of Physics, University of Toronto,
Toronto, Ontario, M5S 1A7, Canada}

\affil{$^2$ Department of Physics and Astronomy, McMaster University, Hamilton, Ontario, L8S 4M1, Canada}

\affil{$^*$Author to whom any correspondence should be addressed.}

\email{stephen.julian@utoronto.ca}

\keywords{kagome metal, unconventional transport, magnetism}

\begin{abstract}
We report the 
  growth and characterization 
  of single crystals of 
  the disorted-kagome lattice compound 
  CrPdAs. 
Spin-glass behaviour with $T_{SG} \sim 60\ {\rm K}$ 
  is observed in all crystals tested. 
Some growths show in addition 
  a magnetic impurity phase with $T_c$ around 200 K, 
  but annealing produces single-phase crystals 
  without the ferromagnetic impurity phase.
Single-phase crystals nevertheless have $29\pm 5\%$ anti-site 
  disorder of the Cr and Pd sites, 
  similar to a previous generation of 
  flux-grown polycrystalline samples. 
We observe 
  a large linear-coefficient of the heat capacity at low temperature, 
  $\gamma = 23 \pm 3$ mJ/mole\,K$^2$, 
  which is typical of kagome metals. 
The calculated band structure shows 
  several Dirac band-crossings very near $E_F$ whose degeneracy is 
  lifted when spin-orbit interaction is included. 
Our most curious finding is a `non-metallic' in-plane resistivity, 
  extending over the entire measured temperature range from 
  300 K down to 2 K.
This resistivity 
  is quasi-linear below about 130 K, 
  and
  shows no sign of saturation down to 
  the lowest temperature measured. 

\end{abstract}

\section{Introduction}
\label{sec-intro}

The TT$'$X family of materials, 
  where T and T$'$ are transition metals 
  and X is arsenic or phosphorus, 
  were extensively 
  investigated by R.\ Fruchart and collaborators 
  in the 1970's and 80's  
  \cite{Fruchart1982}.
Attention at that time focused on magnetic properties, 
  while 
  other thermodynamic and transport properties were 
  relatively neglected.

\begin{figure}
\begin{center}
\includegraphics[width=12cm]{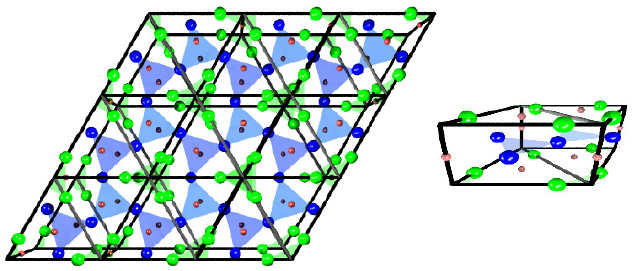}
\end{center}
\caption{Left: A view, looking down the $c$-axis, of 
  the crystal structure of CrPdAs, a TT$'$X compound with the 
  P$\overline{6}$2m structure.  
  3 by 3 unit cells are shown.
Right: a single unit cell, viewed from the side. 
  The Cr sites (blue balls) occupy the vertices of a 
    distorted kagome lattice (blue triangles). 
  Pd (green) form trimers. As atoms are shown as pink balls.
  Note that in CrPdAs 
  there is considerable Cr-Pd antisite disorder (see text).} 
\label{fig:xtalstructure}
\end{figure}

Recently, a study of the magnetic couplings derived from electronic structure 
  calculations on CrRhAs \cite{Huang2023} 
  pointed out that TT$'$X compounds with the P$\overline{6}$2m 
  structure can be classified as kagome metals, 
  in which frustrated hopping paths on 
  the kagome lattice 
  \cite{Mielke1991,Ohgushi2000,Lacroix2009,Kida2011} 
  can create quasi-localized electrons. 
 Kagome metals have received considerable attention recently 
  not least for topological band structure effects, 
  including a pronounced anomalous Hall effect and strange metallic 
  behaviour \cite{Ye2018,Ye2023}. 
The distorted kagome lattice on one of the transition metal sites
  in the TT$'$X structure 
  can be seen in Fig.\ \ref{fig:xtalstructure}. 

Earlier, signs of 
  unusual transport properties in P$\overline{6}$2m TT$'$X systems 
  came from the observation that the 
  electrical resistivity of 
  FeCrAs \cite{Wu2009}, 
  which is in the metallic range,  
  rises continuously from 800 K down to millikelvin 
  temperatures. 
This was termed `non-metallic' resistivity, 
  as opposed to `insulating behaviour', 
  because there is no gap in the density of states: 
  the resistivity at low temperature fits a sub-linear power law, 
  and heat capacity measurements show  
  a high density of states at $E_F$, with Sommerfeld 
  coefficient $\gamma = 32 {\rm\ mJ/mol\,K^2}$.

Very recently, quasi-linear `strange metal' resistivity 
  (but with a positive temperature coefficient) and a large 
  anomalous Hall effect have been 
  reported near a pressure-induced magnetic quantum critical point 
  in CrNiAs \cite{Shen2025}, 
  another TT$'$X-arsenide with the P$\overline{6}$2m structure. 

CrPdAs is also a member of this family. 
Earlier studies of CrPdAs used polycrystalline samples 
  grown by solid-state reaction.  
It was reported that 
  CrPdAs is a spin-glass with a freezing temperature 
  around 35 K \cite{Kanomata1991,Kaneko1992}. 
The spin-glass behaviour was attributed to a considerable degree 
  (approximately 30\%) of 
  Cr and Pd anti-site disorder \cite{RoyMontreuil1979}. 
Other transport and thermodynamic properties were not reported. 

Here we report the growth of single crystals of 
  CrPdAs, together with measurements of elastic powder x-ray and 
  neutron diffraction, as well as the  
  resistivity, heat capacity, and magnetic properties. 
We find spin-glass behaviour but 
  with a higher freezing temperature than the earlier study,  
  $T_F \sim 60$ K,
  despite similar levels of site disorder. 
We find that the linear coefficient of heat capacity is 
  $\gamma = 23\pm3$ mJ/mole\,K$^2$, which is 
  large for a transition metal compound,  
  but it is 
  typical of kagome metals (e.g.\ \cite{Ishii2012,Ortiz2019,He2014}) 
  and similar to FeCrAs \cite{Wu2009}. 
We have carried out LDA band structure calculations 
  (for the experimentally determined structure but without site disorder) 
  and find multiple Fermi surfaces and a high density of 
  states, with flat bands at the top of the Brillouin zone, 
  produced by multiple 
  Dirac-type band crossings very close to the Fermi energy, which are gapped 
  when the spin-orbit interaction is  taken into account. 
Our most interesting finding is that the in-plane resistivity, 
  $\rho_{ab}(T)$,  
  is `non-metallic' over the entire temperature range measured, rising 
  monotonically with decreasing temperature from 
  room temperature to 2 K, with no sign of saturation down to 2 K.
Intriguingly, $\rho_{ab}(T)$ is quasi-linear below about 
  130 K. 
$\rho_c$, in contrast, is only weakly temperature dependent. 

\section{Crystal Growth}
\label{sec-Xtals}

Polycrystalline CrPdAs 
  has previously been grown by solid state diffusion, in sealed quartz ampoules
  \cite{RoyMontreuil1979}. 
The existence of a low-melting-point ($\sim 620^\circ$C) 
  eutectic of As and Pd is a complication in this method. 

We have grown single crystals by slow cooling of 
  the stoichiometric melt, a 
  method developed for FeCrAs \cite{Wu2009}. 
We carried out three growths, which we refer to below 
  as G1, G2 and G3.  
In all three growths, 
  stoichiometric amounts of Cr, Pd and As were placed in an 
  alumina crucible and sealed in an evacuated quartz tube, 
  and then heated in a box furnace.
As explained below, two of the growths, which had a maximum 
  temperature of 1100$^\circ$C, and no annealing step, 
  show signs ferromagnetic impurity phases, 
  but the third growth, G3, for which we carried out an additional 
  annealing step, is single phase and 
  shows no ferromagnetic anomalies. 

\begin{figure}
\begin{center}
\includegraphics[width=12cm]{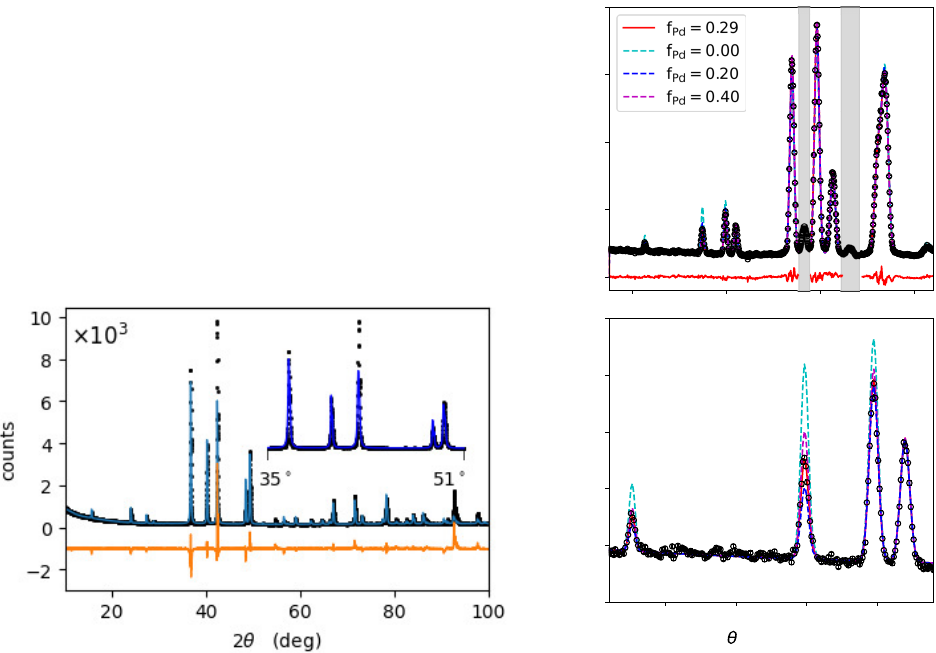}
\end{center}
\caption{(a) Powder x-ray diffraction on growth G3, fitted with 
  Profex \cite{Dobelin2015}. The inset zooms in on the strong peaks 
  between $2\theta = 35$ and 51 degrees. 
 There is no sign of secondary 
 phases (cf.\ sample G1, Appendix B). 
  Despite allowing for rather large grain anisotropy, and also Cr/Pd site 
 disorder, the amplitude of some peaks is not well fitted.
(b)
 Powder neutron diffraction spectrum, fitted by varying the
  level of anti-site disorder on the Cr and Pd sites. $\rm f_{Pd}$ is the
  fraction of Pd on Cr sites (and vice-versa). The best fit
  yields ${\rm f_{Pd}} = 0.29\pm0.05$\%.
The peaks at $\sim 56^\circ$ and $\sim 66^\circ$ highlighted in grey arise
  from the aluminum sample cans and are excluded from refinement. 
(c) The (001) peak is particularly
  sensitive to anti-site disorder.}
\label{fig:PXR}
\end{figure}



For the G3 growth, we melted the sample three times in all. 
The first heating sequence was: 
  30\C\htime{24}600\C\htime{60}600\C\htime{35}1100\C, power off. 
The resulting ingot was removed from the quartz tube, 
  turned over and 
  transferred to a fresh crucible and quartz tube, 
  and then remelted.  
The second sequence was: 
  30\C\htime{30}1100\C, power off. 
The ingot was again transferred to a fresh crucible and quartz tube. 
A third melt had sequence: 
  30\C\htime{30}1150\C\htime{5}1150\C\htime{72}950\C\htime{72}950$^\circ$C, 
  power off.
The 72 hour dwell at 950$^\circ$C was intended as an annealing step. 

Note that slow heating to 600$^\circ$C on the first run, and subsequent 
  dwell for 60 hours at 600$^\circ$C, is for safety reasons: 
the vapor pressure of arsenic reaches 1 atmosphere at 614$^\circ$C. 
The temperature therefore is held just below this point for a long time to 
  make sure that the arsenic vapor has reacted with the transition metals, 
  after which the vapor pressure is low enough that the quartz tube will 
  not crack upon subsequent heating.
As noted above, PdAs has a eutectic at 620$^\circ$C, so this procedure 
  may not be necessary in CrPdAs, but we did not risk dispensing with this 
  step. 

The ingots, when removed from the crucibles, have cracks in them. 
Carefully separating the pieces of ingot along these crack boundaries 
  yields some mm-sized single crystals that are irregularly shaped (i.e.\ 
  they do not have facets) 
  as we also found in FeCrAs \cite{Wu2009}.

Fig.\ \ref{fig:PXR}a shows the powder x-ray diffraction pattern of 
  crushed crystals from growth G3. 
It shows no sign of a second phase, which is significant 
  given that this growth, 
  unlike G1 and G2, 
  shows no sign of ferromagnetic impurity phases 
  (see below). 
In Appendix B the powder x-ray spectrum from growth G1 is compared 
  with the spectrum of G3 shown in Fig.\ \ref{fig:PXR}a.

In order to test for anti-site disorder, which was reported in 
  \cite{RoyMontreuil1979}, we carried out powder neutron diffraction at 
  the McMaster All-purpose Diffractometer (MAD),  
  at the McMaster Nuclear Reactor. 
An incident neutron wavelength of 2.21Å was used.
The results are shown in Fig.\ \ref{fig:PXR}b and \ref{fig:PXR}c. 
Fitting with varying amounts of site disorder gives a clear best-fit 
  for $\rm f_{Pd}$, the fraction of Cr and Pd on the `wrong' site, 
  of $0.29\pm0.05$.
This is in good agreement with the 0.3 fraction of anti-site disorder 
  reported by \cite{RoyMontreuil1979}. 

We thus conclude that our method yields single-phase CrPdAs, but 
  that there is a similar amount of anti-site disorder compared to 
  earlier polycrystalline samples 
  grown at lower temperature 
  by solid-state reaction.

\section{Heat capacity and magnetization}
\label{sec-TD}

\begin{figure}
\begin{center}
\includegraphics[width=12cm]{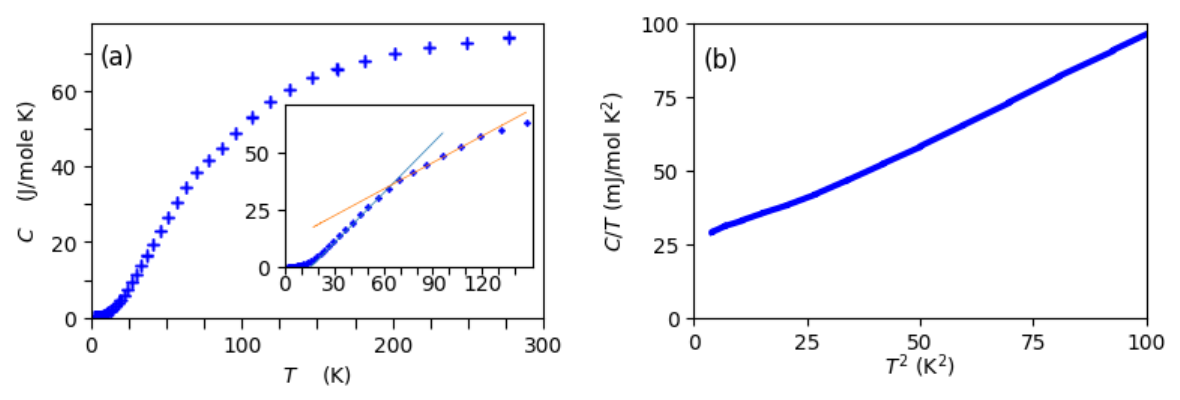}
\end{center}
\caption{
  (a) Heat capacity vs temperature from room temperature to 2 K, 
      measured on a sample from growth G3. 
      The inset focuses on the region below 150 K, and illustrates 
      enhancement of $C(T)$ around 60 K. 
  (b) $C/T$ vs.\ $T^2$ from 10 K to 2 K from the same data as (a).}
\label{fig:C_vs_T}
\end{figure}

Fig.\ \ref{fig:C_vs_T}a  
  shows heat capacity vs.\ temperature from room temperature to 2 K, 
  measured on a PPMS system using the relaxation method.
$C(T)$ agreed between the different growths within a few percent. 
Sample masses varied from 15 to 75 mg. 
The heat capacity is relatively featureless, but there is 
  a broad region of enhanced heat capacity centred around 60 K (see inset), 
  which we show below is the freezing temperature for a spin-glass phase. 
Fig.\ \ref{fig:C_vs_T}b 
  shows $C/T$ vs.\ $T^2$ in the $T\rightarrow 0$ K limit.
There is a slight change of slope around $T^2 = 25$ K$^2$. 
The linear coefficient of heat capacity extrapolates 
  to 20 mJ/mole\,K$^2$ using the high temperature region, 
  and 26 mJ/mol\,K$^2$ using the low temperature region, 
  so we set $\gamma = 23 \pm 3$ mJ/mole\,K$^2$. 
This value of $\gamma$ is large for a 
  transition metal compound, but is not uncommon in 
  kagome metals.  
The $T^3$ coefficient is 
  $\beta = (0.74 \pm .02)$ mJ/(mol\,K$^4$), 
  corresponding to a Debye temperature of $200\pm 2$ K. 

\begin{figure}
\begin{center}
\includegraphics[width=12cm]{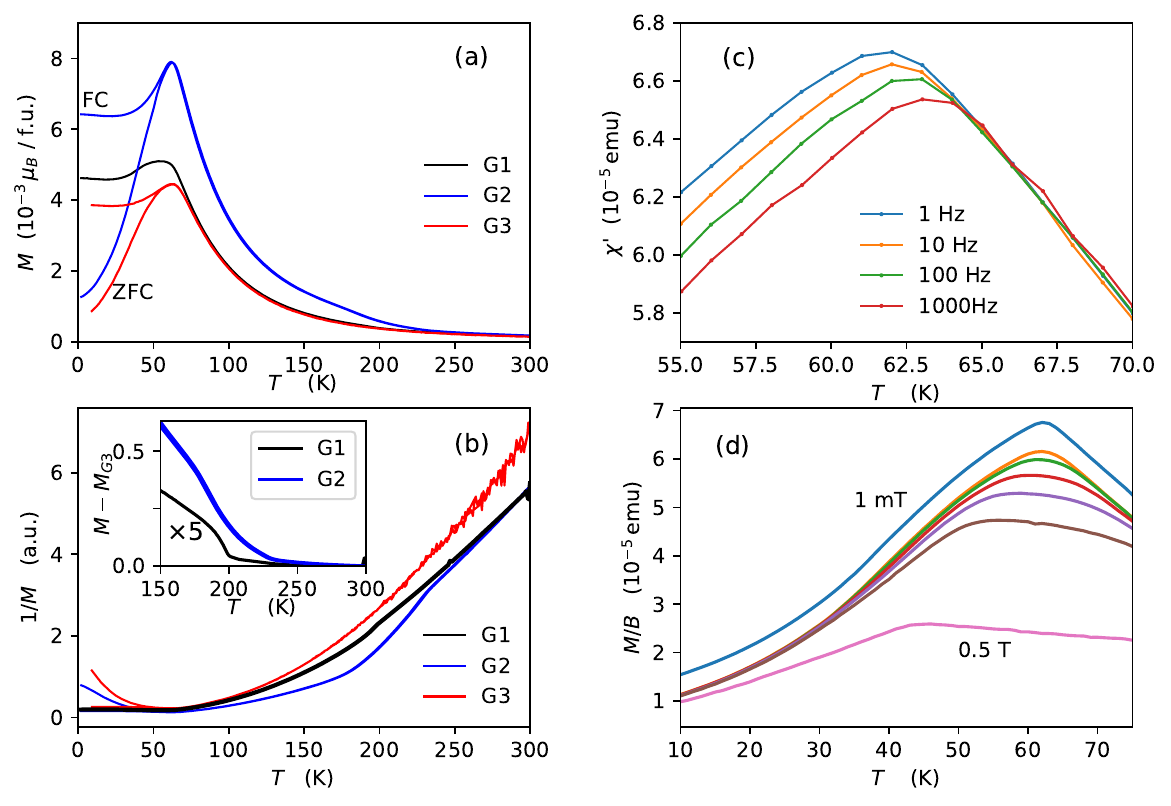}
\end{center}
\caption{
  Magnetization of CrPdAs. 
  (a) Magnetization vs.\ temperature for polycrystalline samples 
    from our three 
    growth runs, G1, G2 and G3. 
    FC refers to field cooling in a field of 100 Oe, while 
    ZFC refers to samples coolied in zero field, prior to application 
    of an external field of 100 Oe. 
    Sample G1 was measured only in the FC condition. 
    The separation of FC and ZFC curves below 60 K is a possible signature 
      of spin-glass behaviour. 
  (b) $1/M$ vs.\ $T$ shows that there are ferromagnetic impurity phases in 
     the G1 and G2 samples, with an enhancement of magnetization setting in 
     over a broad temperature range (230 to 190 K) in sample G2, and more 
     weakly over a narrow temperature range centred on 200 K in G1. 
     This is seen more clearly in the inset, which shows the magnetization of 
     growths G1 and G2 after subtraction of the smoothly-varying magnetization 
     of the G3 sample.
  (c) Real part of the ac-susceptibility, $\chi'$, vs.\ T 
      from 1 Hz to 1000 Hz. 
      The observed shift of the peak to higher temperatures with increasing 
      ac-frequency 
      is a hallmark of spin-glass behaviour. 
  (d) $M/B$, 
      measured at 10 Hz, vs.\ T at applied 
      fields of (top to bottom) 1 mT, 5 mT, 10 mT, 25 mT, 50 mT, 0.1 T and 
      0.5 T. The spin-glass cusp shifts to lower temperature with increasing 
      field. 
} 
\label{fig:MvsT}
\end{figure}

Magnetic properties were measured on polycrystalline material 
  on an MPMS system.
The results are shown in Fig.\ \ref{fig:MvsT}. 
We see a pronounced difference between 
  the three growths. 

Fig.\ \ref{fig:MvsT}a shows $M$ vs.\ $T$, measured in a field of 100 Oe, 
  for crystals from the three 
  growths.  
For G2 and G3 we show both field-cooled (FC) and zero-field-cooled 
  (ZFC) conditions. 
G1 was measured only in the FC condition. 
The results differ markedly at low temperature, with sample G2 having nearly 
  twice the induced magnetization of sample G3. 
The lower plot, Fig.\ \ref{fig:MvsT}b shows the cause of this 
  discrepancy, which is ferromagnetic impurity phases in samples G1 and G2. 

The main plot in Fig.\ \ref{fig:MvsT}b shows $1/M$, which serves to 
  emphasize the behaviour at high temperature where $M$ is small.
At high temperature there is an offset between the three curves: 
  G1 and G2 agree, but G3 
  has a lower magnetization (higher $1/M$). 
More importantly, 
  the G1 and G2 crystals show suddden slope changes 
  in the region around 200 K, 
  while the slope of G3 evolves smoothly. 
The inset of Fig.\ \ref{fig:MvsT}b shows the magnetization of the 
  G1 and G2 crystals after subtraction of a polynomial fit to the 
  magnetization of the G3 crystal. 
Starting above 250 K the curves begin to deviate from zero, showing an 
  additional ferromagnetic contribution in G1 and G2.  
In sample G1 this is quite weak (the G1 curve in Fig.\ \ref{fig:MvsT}b 
  has been multiplied by 5), but there is a clear onset of a 
  ferromagnetic phase at 200 K, signaled by the sudden upturn in the 
  curve. 
In sample G2 the ferromagnetic contribution is much stronger, and seems 
  to have a range of $T_C$'s, given the absence of a well-defined break 
  in slope such as we see in the G1 curve. 

We conclude that the difference in maximum moment, at $\sim 60$ K, 
  is due to ferromagnetic impurity phases in samples G1 and G2. 
We therefore take sample G3 to be the intrinsic behaviour.
We have not been able, so far, to identify the ferromagnetic impurity.
In Appendix B we show that the G1 powder x-ray spectrum shows signs of 
  a possibly orthorhombic impurity phase as well as some phase separation 
  of the dominant hexagonal phase. 

All samples show a peak in the induced magnetization 
  at $\sim 60$ K, below which 
  the field-cooled (FC) and zero-field-cooled 
  (ZFC) lines deviate from each other. 
Differences between FC and ZFC magnetization 
  can be an indication of spin-glass behaviour, which was previously 
  found in CrPdAs below about 35 K in polycrystalline samples 
  \cite{Kanomata1991,Kaneko1992}. 

A more rigorous test of spin-glass behaviour is the frequency dependence 
   of the 
  magnetic susceptibility. 
Fig.\ \ref{fig:MvsT}c shows how the peak in the 
  ac-susceptibility vs.\ temperature changes as 
  the frequency of the applied ac-field is varied, 
  while Fig.\ \ref{fig:MvsT}d shows how the ac-susceptibility vs.\ 
  temperature evolves 
  as a function of applied $dc$-field,  
  at a fixed measurement frequency of 10 Hz. 
The upward shift of the maximum in $\chi'$ with measurement frequency 
  in particular is an accepted signature of 
  spin-glass behaviour (e.g.\ \cite{Mulder1982}). 
Despite the similar 
  level of site disorder 
  our samples have 
  a signficantly higher spin-glass temperature than 
  was reported for samples grown by solid-state reaction
  \cite{Kanomata1991}, 
  but some of this discrepancy may be due to measurement conditions. 
The lower plot (Fig.\ \ref{fig:MvsT}d) shows that the peak in $M/B$ 
  is rapidly suppressed with field, falling to about 43 K by 0.5 T.
Reference \cite{Kanomata1991} used a field of 0.2 T.  

\section{Band Structure}
\label{sec-bs}

\begin{figure}
\begin{center}
\includegraphics[width=14cm]{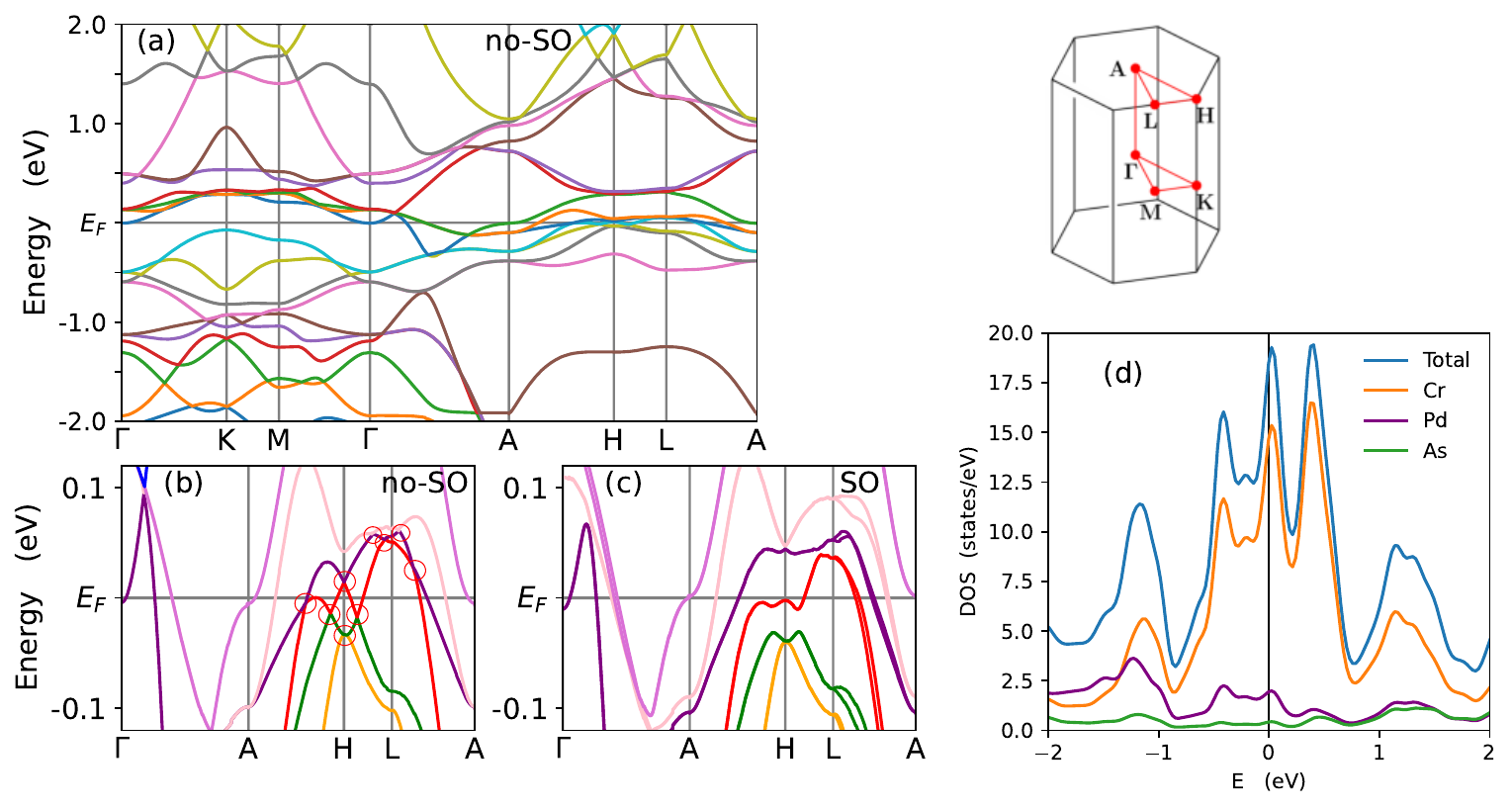}
\end{center}
\caption{
  (a) Spaghetti plot of the calculated band structure of CrPdAs, 
      without spin-orbit coupling, 
      plotted from -2.0 eV to 2.0 eV, relative to $E_F$.  
      The Brillouin zone is shown on the right. 
  (b) The same band structure as in (a), but zooming in on the 
        region within  
        $\pm 0.1$ eV of $E_F$. 
        Across the top of the Brillouin zone (surrounding the H point) there 
            are several band crossings very close to $E_F$ (circled). 
  (c) Same conditions as in (b), but for 
      the band structure with spin-orbit coupling included. 
      This lifts the degeneracy at the band crossings, and produces a flat 
      band very close to $E_F$ around the H-point. 
  (d) The density of states from the calculated band structure. 
  Top to bottom: total, chromium, palladium, and arsenic contributions to the 
  density of states.
 }
\label{fig:Spag}
\end{figure}

We have calculated the band structure of CrPdAs 
  using WIEN2k \cite{Blaha2020}. 
We assumed no site disorder, 
  and performed the calculation both without spin-orbit coupling 
  (Fig.\ \ref{fig:Spag}a,b), and 
  with spin-orbit coupling (Fig.\ \ref{fig:Spag}c). 

Across the top of the Brillouin zone there are several band crossings 
  within 30 meV of $E_F$ (circled in Fig.\ \ref{fig:Spag}b). 
Typically, when spin-orbit coupling lifts the degeneracy at a Dirac 
  point one obtains ``massive Dirac fermions", but in this case the 
  density of Dirac crossings is so high that 
  the result is better described as a 
  flat band sitting within a few meV of $E_F$ (see in Fig.\ \ref{fig:Spag}c, 
  the red band just below $E_F$, in a region centred on the H-point).

Fig.\ \ref{fig:Spag}d shows the calculated density of states. 
The density of states at $E_F$, dominated by Cr $d$-electrons,
is a result of moderately flat bands: 
  the calculated Sommerfeld coefficient is 15 mJ/mole\,K$^2$, which is 
  more than half of the measured $\gamma = 23 \pm 3$ mJ/mole\,K$^2$.
The mass enhancement due to many-body interactions is therefore modest.

\section{Resistivity}
\label{sec-RvsTH}

\begin{figure}
\begin{center}
\includegraphics[width=8cm]{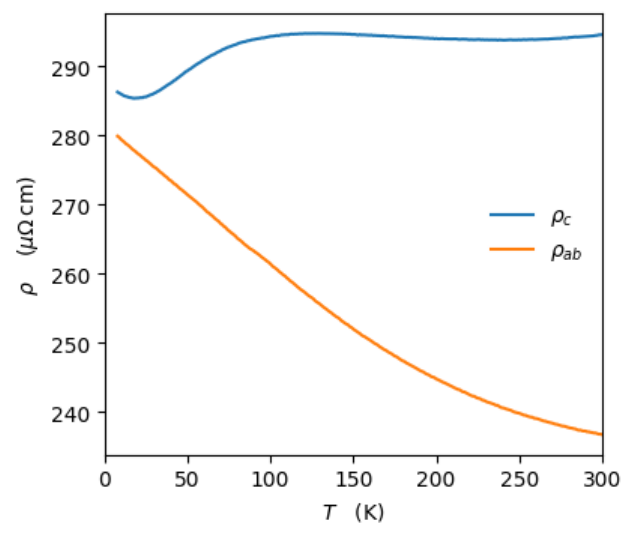}
\end{center}
\caption{Resistivity vs.\ temperature for CrPdAs, for crystals from 
  growth G3. 
  The $c$-axis resistivity (upper, blue curve) 
    is relatively weakly temperature dependent, 
  and falls (i.e.\ is `metallic') across the spin-glass transition 
  at 60 K. 
  The $ab$-plane resistivity (lower, orange curve) 
    is non-metallic across the entire 
    temperature range, and is quasi-linear below about 130 K, with no 
    apparent change in slope through the spin-glass transition, and 
    no apparent saturation at low temperature.  In Appendix B we 
    show that 
    the in-plane resistivity of crystals 
    from growths G1 and G3 are 
    consistent with each other; 
    the resistivity of growth G2 was not measured.}
\label{fig:rhovsT} 
\end{figure}

Fig.\ \ref{fig:rhovsT} shows resistivity vs.\ temperature measured 
  on single crystals from growth G3 with the current 
  along the $c$-axis ($\rho_c$) and in-plane ($\rho_{ab}$). 
The measurements were done on a PPMS system.  
A plot comparing the resistivity of G3 with G1 can be found in Appendix B. 
The in-plane resistivities agree well.
$\rho_c(T)$ in contrast shows some sample dependence but,  
  as discussed in Appendix B, 
  it was difficult to measure $\rho_c$ without contamination from 
  $\rho_{ab}$ whose $T$ dependence is much stronger. 

The intrinsic $c$-axis resistivity is nearly temperature independent 
  from room temperature down to $\sim 100$ K, 
  then  
  falls slightly between 90 K and 30 K, 
  then shows an upturn in 
  the low temperature limit. 
There is some resemblance to 
  the $c$-axis resistivity of FeCrAs \cite{Wu2009,McGuire2025} 
  which, although it has a stronger temperature dependence, 
  is non-metallic down 
  to the N\'eel temperature $T_N \simeq 120$ K, 
  below which there is a region of metallic resistivity, 
  followed below about 25 K by a return to non-metallic resistivity. 

The in-plane resistivity
  is perhaps the most curious of our results on CrPdAs: a 
  non-metallic resistivity extending over the entire measured temperature 
  range, with quasi-linear behaviour from about 130 K down to the 
  lowest temperature measured, 
  with no anomaly associated with the spin glass transition. 
We note that the leads must be carefully aligned in this measurement in 
  order to avoid contamination of 
  the in-plane resistivity with the non-linear c-axis resistivity. 

The non-metallic in-plane resistivity of CrPdAs is again reminiscent of 
  the in-plane resistivity of FeCrAs \cite{Wu2009}, 
  which is also non-metallic over the entire range measured, 
  but there are differences. 
Firstly, in CrPdAs the absolute change in resistivity is 
  weaker 
  than in FeCrAs, with $\rho(2{\rm\ K})-\rho(300 {\rm\ K})$ being about 
  50 $\mu\Omega\,{\rm cm}$ in CrPdAs, 
  vs. $\sim180\,\mu\Omega\,{\rm cm}$ in FeCrAs.
Secondly, however, there is the above-noted linearity of the 
  non-metallic resistivity of CrPdAs over a large temperature range that, 
  interestingly, seems unaffected by the spin-glass transition. 
This behaviour was not seen in FeCrAs. 
Writing the resistivity as $\rho(T) = \rho_\circ + AT^n$, 
  fitting over various temperature ranges with a lower bound of 
  2 K and an upper bound between 60 K to 
  130 K gives quite consistent results: 
  $\rho_\circ = 282 \pm 0.5 \,\mu\Omega\,{\rm cm}$, 
  $A = -0.24\pm 0.1 \,\mu\Omega\,{\rm cm/K}^n$, and 
  $n = 1.00 \pm 0.03$. 
 
The significance, if any, of the quasi-linearity is difficult to 
  judge. 
A linear {\it positive} slope in the resistivity is 
  of course common in metallic conductors. 
At high temperature 
  it is well-understood to arise from electron-phonon 
  scattering.  
Linear metallic resistivity extending below 10 K is rare, 
  but it has been seen in strongly correlated systems near quantum 
  critical points, and in optimally-doped cuprates which may also be near 
  to a quantum critical point \cite{Ramshaw2015}), in which case it is 
  often 
  connected to ``strange metallic behaviour" (see e.g.\ \cite{Phillips2022}), 
  and ``Planckian dissipation" (see e.g.\ \cite{Zaanen2004,Bruin2013}). 

In CrPdAs, the amplitude of the linear resistivity does approximately 
  match the estimated Planckian limit, as shown in Appendix A,  
  except that of course it has a negative $A$ coefficient.  

Quasi-linear resistivity with a negative temperature coefficient 
  was observed in doped Kondo alloys such as Y$_{1-x}$U$_{x}$Pd$_3$ 
  \cite{Seaman1991,Maple1995}. 
In that case, however, it was accompanied by a 
  highly enhanced and logarithmically divergent 
  heat capacity in the $T\rightarrow 0$ K limit, 
  which we do not observe in CrPdAs. 
Nevertheless, one of the models developed for this phenomenon, including 
  disordered Kondo lattice \cite{Miranda1996}, 
  or Griffiths phase physics \cite{CastroNeto1998}, 
  may be applicable here. 

The most interesting aspect of $\rho_{ab}$ may be that the 
  quasi-linear behaviour 
  extends down to at least 2 K without saturating.
The usual mechanism for non-Fermi-liquid behaviour extending 
  to low temperatures is a quantum critical point, 
  but the magnetization and heat capacity do not suggest that 
  CrPdAs is near a quantum critical point. 
Exponential suppression of the Kondo temperature by Hunds coupling 
  \cite{Nevidomskyy2009} may allow Kondo effects to extend 
  to low temperature, but there is no particular reason why the 
  resistivity should depend linearly on $T$ in such a model. 

Finally, we note that 
resistivity vs.\ temperature has previously been measured in 
  another kagome system with a spin-glass transition, in
  Co$_{1-x}$Fe$_{x}$Sn for $x = 0.08$ and $x = 0.17$
  \cite{Sales2021}.
In that case the resistivity is metallic, but 
  also anomalous:
  the temperature dependence is sub-linear over the entire temperature
  range measured ($\sim 1$ to 300 K), 
  and the resistivity does not saturate down to the lowest
  temperature measured.
As in CrPdAs there is no anomaly in the resistivity at the spin-glass 
  transition.
The authors of Ref.\ [30] 
  note that the sub-linear $T$-dependence could be due to a 
  peak in the density of states lying close to $E_F$.

\section{Conclusions}
\label{sec-Concl}

We have grown single crystals of the distorted-kagome-lattice 
  compound CrPdAs. 
We find 
  Cr-Pd anti-site disorder that is similar in magnitude, $\sim 30\%$, 
  to a previous 
  generation of polycrystalline samples grown 
  at lower temperature by solid-state diffusion. 

Like FeCrAs, CrPdAs has intriguing transport properties, notably  
  non-metallic resistivity over a large temperature range, 
  with a non-saturation of $\rho_{ab}$ down to at least 2 K. 
Interestingly, the (negative) slope of $\rho_{ab}$ is 
  quasi-linear from $\sim 130$ K down to 2 K, 
  and seemingly unaffected by spin-glass ordering around 60 K.  
The $c$-axis resistivity is only weakly temperature dependent. 

The linear coefficient of the heat capacity 
  is large for a transition metal, with $\gamma = 23 \pm 3$ 
  mJ/mol\,K$^2$, but in line with other kagome metals. 
Similarly, the band-structure, like some other kagome metals, has several Dirac 
  band-crossings very close to $E_F$. 
When spin-orbit is included in the band-structure calculation the 
  degeneracy is lifted to give a flat band just below $E_F$. 

\funding 

This research was supported by the
  Natural Sciences and Engineering Research Council of Canada
  (NSERC RGPIN-2019-06446). 
Use of the MAD beamline at the McMaster Nuclear Reactor is 
  supported by McMaster University and the Canada Foundation for Innovation.

\data

The authors confirm that the data supporting the findings of this study 
  are available within the article and Appendix B. 

\roles

Benny Lau: Investigation (equal), Data curation (equal), 
  Formal analysis (equal), 
  Writing -- First Draft (lead) 

\vspace{12pt}

\noindent
Wenlong Wu: Conceptualization (supporting), 
  Investigation (equal), Investigation (equal), Methodology (equal)

\vspace{12pt}

\noindent
Bo Yuan: Investigation (equal), Formal analysis (equal), Writing (supporting)

\vspace{12pt}

\noindent
Julian Nickel: Investigation (equal), Formal analysis (supporting)

\vspace{12pt}

\noindent
Stephen Julian: Conceptualization (lead), Formal analysis (equal), 
  Funding acquisition (lead), Investigation (equal), 
  Project Administration (lead),
  Supervision (lead), Writing -- First draft (supporting), Writing -- review \& editing (lead)

\section{Appendix A: Estimate of the Planckian resistivity}
\label{sec-A1}

Linear resistivity, with a positive temperatue coefficient, 
  extending to low temperature has 
  been connected to so-called Planckian dissipation. 
It is perhaps of interest to 
  compare the magnitude of the change in $\rho_{ab}(T)$ 
  with the predictions of Planckian dissipation: 
  we show here that they are of comparable magnitude, 
  but of course $d\rho_{ab}/dT$ of CrPdAs is in the `wrong' direction, 
  with our $\rho_{ab}(T)$ having a negative temperature coefficient. 

The Planckian resistivity is estimated from the Drude resistivity:
\begin{eqnarray}
\sigma = \frac{n{\rm e}^2 \tau}{m^*},
\end{eqnarray}
with the substitution $\tau = \hbar/k_BT$. Thus
\begin{eqnarray}
\rho = 1/\sigma = \frac{m^*k_BT}{n{\rm e}^2\hbar}, 
\end{eqnarray}
i.e.\ we obtain linear-in-$T$ resistivity, $\Delta\rho = AT^1$, with 
  $A = m^*k_B/n{\rm e}^2\hbar$. 

For CrPdAs we can obtain $n$ from band-structure calculations: 
  $n = 2.2\times 10^{28} {\rm m^{-3}}$. 
For the effective mass, 
  our band structure calculation 
  gives $m^*$ in the range of 5 to 10 $m_e$.
If we use the enhancement of about 2 for the measured vs.\ the 
  calculated $\gamma$, then  
  $m^*$ would be between $10m_e$ and $20m_e$. 
(There may be much heavier masses around the H-point, but there 
  is no reason to assume that they will dominate the conductivity.) 
This gives
\begin{eqnarray}
A   &=& (2.1 {\rm\ to\ } 4.2)\times 10^{-1} \mu\Omega\,{\rm cm/K}, 
\end{eqnarray}
Going from 0 to 100 K we would thus expect a change of 
  20 to 40 $\mu\Omega\,{\rm cm}$, 
  which overlaps our observed change of just over 
  20 $\mu\Omega\,{\rm cm}$.

\section{Appendix B: Comparison of the resistivity and x-ray spectra of growths 1 and 3}
\label{sec-A2}

\begin{figure}
\begin{center}
\includegraphics[width=14cm]{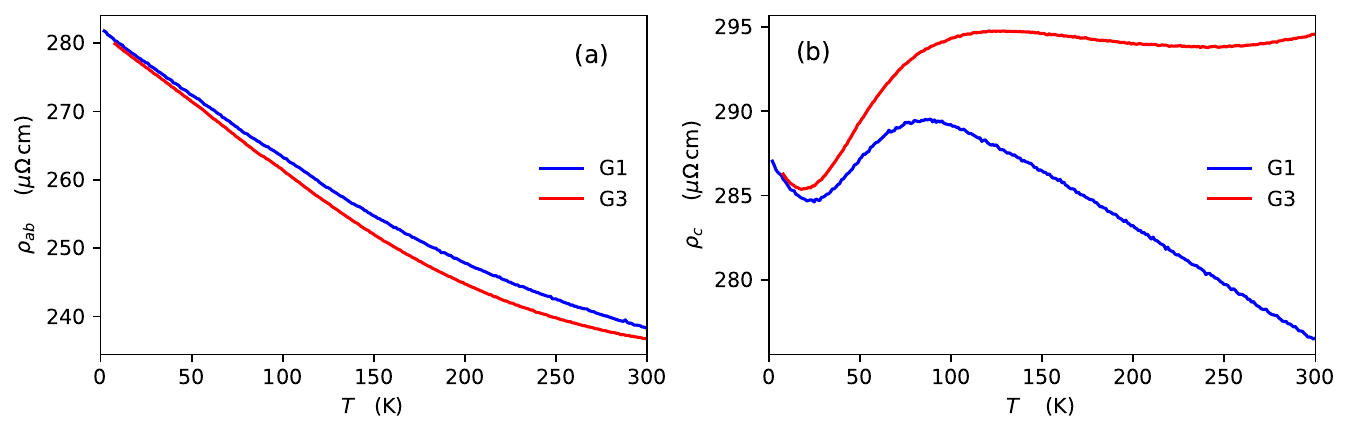}
\end{center}
\caption{Comparison of $\rho(T)$ for crystals from growths G1 and G3. 
(a) $\rho_{ab}(T)$, (b) $\rho_{c}(T)$. 
Note the larger $y$-axis scale in (a), reflecting the 
  stronger $T$ dependence of $\rho_{ab}(T)$. 
In each plot the data have been shifted vertically by a few percent 
  so that they agree at low temperature. 
The `G3' data is the same as in Fig.\ \ref{fig:rhovsT}. }
\label{fig:rho_compare}
\end{figure}

We have measured $\rho_{ab}(T)$ and $\rho_c(T)$ on several different crystals 
  from growths G1 and G3. 
The in-plane resistivity, 
  $\rho_{ab}$, 
  shows good consistency between crystals from G1 and G3, 
  as illustrated in Fig.\ \ref{fig:rho_compare}a. 
Data from the G1 sample has been rescaled by approximately 10\% 
  in Fig.\ \ref{fig:rho_compare}a 
  so that the resistivities agree at low temperature: 
  the difference in absolute magnitude is probably 
  due to uncertainty in lead placement, but may involve some actual
  sample-dependence. 
The quasi-linear dependence of $\rho_{ab}$ upon temperature $T$ 
  is seen at low temperature in both samples. 

Our measured $\rho_{c}$, in contrast, shows some sample dependence 
   (see Fig.\ \ref{fig:rho_compare}b). 
The dip 
   centred on about 60 K in Fig.\ \ref{fig:rhovsT}a was seen in 
   all samples, but 
   there were differences in the overall slope, which may be due 
   to contamination by the much-stronger $T$ dependence of 
   $\rho_{ab}(T)$. 
The data that we have selected as `intrinsic', and which we showed in 
  Fig.\ \ref{fig:rhovsT}, has the least-negative slope at high temperature, 
  which we take to be the least-contaminated by in-plane resistivity. 

For the resistivity of the G1 sample, the onset of ferromagnetism around 
 200 K does not show up in either resistivity curve. 

\begin{figure}
\begin{center}
\includegraphics[width=12cm]{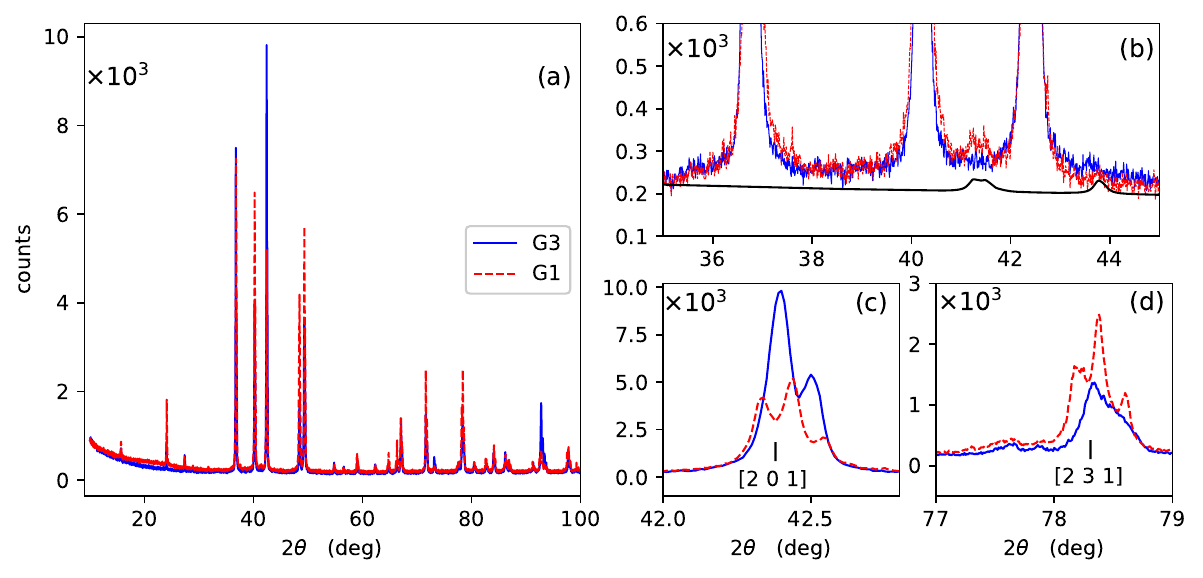}
\end{center}
\caption{Comparison of powder x-ray diffraction patterns of 
  samples from growths G1 (red dashed line) and G3 (blue solid line). 
(a) Spectra over the full measured range of $2\theta$ from 10 to 100 degrees. 
The main peaks match well between the samples, both in position and width. 
Variations in height may be due to intrinsic or extrinsic (e.g.\ grain 
 orientation) effects. 
(b) There is a very weak impurity-phase signature in the G1 spectrum 
  (red line, weak peaks at approximately 41.3 degrees), 
  which we have fitted to an orthorhombic phase (solid black line). 
The data are not 
  sufficiently good to definitively fit the impurity peaks.  
(c) and (d) In addition, the G1 pattern (red dashed line) 
  shows splitting of some of the 
  low-intensity peaks in the spectrum, that are unsplit in 
  the G3 sample (blue solid line).}
\label{fig:PXRcomparison}
\end{figure}

In an attempt to determine the crystal structure of the 
  the magnetic impurity phase, which we postulate to exist in 
 growths G1 and G2, 
 we compare the powder x-ray diffraction patterns of G1 and G3 
 in Fig.\ \ref{fig:PXRcomparison}.

There is good agreement in the position and width of the largest peaks 
  in the spectrum, but looking at low-amplitude features 
  there are two subtle differences. 
Firstly, the G1 spectrum has some very weak peaks that show up 
  around 41.3$^\circ$. 
One possible explanation for these peaks is that there is an orthorhombic 
  (Pnma) 
  impurity phase. (We postulate this because the 111 pnictides are typically 
  either tetragonal, orthorhombic Pnma, or hexagonal P$\overline{6}$2m 
  \cite{Fruchart1982}.) 
The strongest peaks in an impurity orthorhombic phase plausibly fit 
  the very weak peaks in the G1 spectrum of Fig.\ \ref{fig:PXRcomparison}(b),  
  but we have not attempted comprehensively fit these peaks, since none of 
  the other predicted 
  peaks are strong enough to show up in the G1 spectrum. Thus, if these 
  weak peaks arise from the 
  the ferromagnetic impurity phase, its structure must still be 
  regarded as unknown.

The second difference between the powder spectra 
  is that some of the diffraction peaks that are single 
  peaks in the G3 sample are split in the G1 sample (see 
  Figs.\ \ref{fig:PXRcomparison}c and d), suggesting some separation 
  into different hexagonal P$\overline{6}$2m phases, perhaps with slightly 
  different compositions. This would presumably involve separation into Cr 
  rich vs.\ Pd rich compositions, but again we were unable to come up with 
  a satisfactory model. 

Table \ref{tab:lparms} compares the crystal structure refinement results 
  for growths G1 and G3, from x-ray powder diffraction, assuming 
  30\% site disorder on the Cr and Pd sites, and also assuming pure 
  P$\overline{6}$2m structure. 

\begin{table}
\caption{Powder x-ray diffraction structure refinement results
  for powdered material from growths G1 and G3 for the  
   P$\overline{6}$2m structure. Cr is at the Wyckoff position
   3f $(x_{Cr},0,0)$ and Pd is at 3g $(x_{Pd},0,\frac{1}{2})$.
  As1 is at 2d $(\frac{2}{3},\frac{1}{3},\frac{1}{2})$, As2 is at
  1a $(0,0,0)$.}
\begin{center}
\begin{tabular}{lcc}
  & G1  & G3  \\
\hline
$a\,({\rm \AA})$ & 6.5073(2) & 6.49970(12) \\
$c\,({\rm \AA})$ & 3.68999(12) & 3.69049(8) \\
$x_{\rm Cr}$ & 0.6039(4) & 0.6036(5) \\
$x_{\rm Pd}$ & 0.2554(4)        & 0.2562(4) \\
Rwp   &    15.71 & 16.23 \\
Rexp &   5.48 & 5.93 \\
GoF   &    2.87 & 2.74 
\end{tabular}
\label{tab:lparms}
\end{center}
\end{table}

We conclude that there are additional peaks in the powder x-ray spectrum 
  of the G1 sample, but 
  our data are not sufficiently good to determine the structure of an 
 impurity phase. 
A key point, nevertheless, 
  is that the powder x-ray spectra support our hypothesis 
  that G3 is single-phase, while G1 has an impurity 
  phase that is ferromagnetic below about 200 K. 


\end{document}